\documentclass[prl,print,showpacs,superscriptaddress,twocolumn]{revtex4}
\usepackage{eurosym}
\usepackage{amssymb}
\usepackage{graphicx}
\usepackage{dcolumn}
\usepackage{bm}
\usepackage{amsmath}

\begin{document}

\title{Searching for Majorana Fermions in 2D Spin-orbit Coupled Fermi
Superfluids at Finite Temperature}
\author{Ming Gong}
\affiliation{Department of Physics, the University of Texas at Dallas, Richardson, Texas,
75080 USA}
\affiliation{Department of Physics and Astronomy, Washington State University, Pullman,
Washington, 99164 USA}
\author{Gang Chen}
\affiliation{Department of Physics and Astronomy, Washington State University, Pullman,
Washington, 99164 USA}
\affiliation{State Key Laboratory of Quantum Optics and Quantum Optics Devices, College
of Physics and Electronic Engineering, Shanxi University, Taiyuan 030006, P.
R. China}
\author{Suotang Jia}
\affiliation{State Key Laboratory of Quantum Optics and Quantum Optics Devices, College
of Physics and Electronic Engineering, Shanxi University, Taiyuan 030006, P.
R. China}
\author{Chuanwei Zhang}
\thanks{Corresponding author, chuanwei.zhang@utdallas.edu}
\affiliation{Department of Physics, the University of Texas at Dallas, Richardson, Texas,
75080 USA}
\affiliation{Department of Physics and Astronomy, Washington State University, Pullman,
Washington, 99164 USA}

\begin{abstract}
Recent experimental breakthrough in realizing spin-orbit (SO) coupling for
cold atoms has spurred considerable interest in the physics of 2D SO coupled
Fermi superfluids, especially topological Majorana fermions (MFs) which were
predicted to exist at zero temperature. However, it is well known that
long-range superfluid order is destroyed in 2D by the phase fluctuation at
finite temperature and the relevant physics is the
Berezinskii-Kosterlitz-Thouless (BKT) transition. In this Letter, we examine
finite temperature effects on SO coupled Fermi gases and show that finite
temperature is indeed necessary for the observation of MFs. MFs are
topologically protected by a quasiparticle energy gap which is found to be
much larger than the temperature. The restrictions to the parameter region
for the observation of MFs have been obtained.
\end{abstract}

\maketitle

A new research direction in low dimensional condensed matter physics that
attracts much recent attention is the study of two-dimensional (2D)
topological quantum states of matter (e.g., fractional quantum Hall effects,
chiral \textit{p}-wave superfluids/superconductors, \textit{etc.}) that
support exotic quasiparticle excitations (named anyons) with Abelian or
non-Abelian exchange statistics \cite{CN,
DasSarma,Ivanov,Duan,Kitaev,Tewari,Fu,JDS}. For instance, it has been shown
recently that a topological cold atom superfluid may emerge from an ordinary
2D $s$-wave Fermi superfluid in the presence of two additional ingredients:
spin-orbit (SO) coupling and Zeeman field \cite{Zhang}. Such topological
superfluids can host Majorana fermions (MFs), non-Abelian anyons which are
their own antiparticles, and may have potential applications in
fault-tolerant topological quantum computation \cite{CN}.

On the experimental side, 2D degenerate \textit{s}-wave Fermi gases have
been realized using highly anisotropic pancake-shaped trapping potential
\cite{Martiyanov,BF,MF,Martin}. Furthermore, SO coupling and Zeeman field
for cold atoms have been generated in recent pioneer experiments through the
coupling between cold atoms and lasers \cite%
{LYJ1,spielman2,jingzhang,jingzhang2,pan,zwierlein}. It seems therefore that
MFs are tantalizingly close to experimental reach in degenerate Fermi gases.
Inspired by the experimental achievements, there have been extensive
theoretical efforts for understanding the physics of SO coupled Fermi gases,
particularly the mean-field crossover from the Bardeen-Cooper-Schrieffer
(BCS) superfluids to Bose-Einstein condensation (BEC) of molecules \cite%
{JPV1,GM,YZQ,HH,Iskin,Yi,LD,Li,Chen,LJ,ZJN,Liu,Iskin2,Zhou}. While the
mean-field theory works qualitatively well in 3D, it may not yield relevant
physics in 2D at finite temperature. For instance, differing from the
mean-field prediction, there is no long-range superfluid order in 2D at
finite temperature due to phase fluctuations \cite{PCH}. At finite
temperature, the relevant physics is the BKT transition \cite{VLB,KT} with a
characteristic temperature $T_{\text{BKT}}$, below which free
vortex-antivortex (V-AV) pairs are formed spontaneously. When the
temperature is further lowered below another critical temperature $T_{\text{%
Vortex}}$, the system forms a square V-AV lattice \cite{DRN,APY}. Because
MFs only live in the cores of spatially well separated vortices, it is
crucial to study the dependence of $T_{\text{BKT}}$ and $T_{\text{Vortex}}$
on the SO coupling strength in 2D Fermi superfluids.

In this Letter, by taking account of phase fluctuations in the finite
temperature quantum field theory, we investigate the finite temperature
properties of SO coupled degenerate Fermi gases en route to clarifying some
crucial issues for the experimental observation of MFs in such systems. Our
main results are the following:

(I) We show that, quite unexpectedly, the SO coupling reduces both $T_{\text{%
Vortex}}$ and $T_{\text{BKT}}$, despite it enhances the superfluid order
parameter $\Delta $.

(II) In the pseudogap phase ($T>T_{\text{BKT}}$), the Fermi gas lacks phase
coherence. While in the vortex lattice phase ($T<T_{\text{Vortex}}$), the
short distance between neighboring vortices may induce a large tunneling
between vortices that destroys the Majorana zero energy states \cite{Cheng}.
Therefore MFs can be observable only in the cores of naturally present V-AV
pairs \cite{Stone} in the temperature region $T_{\text{Vortex}}<T<T_{\text{%
BKT}}$, instead of intuitively expected zero temperature.

(III) MF in a vortex core is protected by a quasiparticle energy gap $%
\gtrsim \Delta ^{2}/2E_{F}$ ($E_{F}$ is the Fermi energy), which is found to
be much larger than the experimental temperature. Such a large energy gap
greatly reduces the occupation probability of non-topological excited states
in the vortex core due to finite temperature, and ensures the topological
protection of MFs.

(IV) The restrictions to the parameter region for the observation of MFs have been obtained. We
also propose a simple strategy for finding experimental parameters for the
observation of MFs.

We consider a 2D degenerate Fermi gas in the presence of a Rashba type of SO
coupling and a perpendicular Zeeman field. In experiments, 2D degenerate
Fermi gases can be realized using a 1D deep optical lattice with a potential
$V_{0}\sin ^{2}(2\pi z/\alpha _{w})$ along the third dimension, where the
tunneling between different layers is suppressed completely \cite%
{Martiyanov,BF,MF, Martin}. The Rashba SO coupling and Zeeman field can be
realized using adiabatic motion of atoms in laser fields \cite%
{LYJ1,spielman2}. The Hamiltonian for this system can be written as ($\hbar
=K_{B}=1$)
\begin{equation}
H=H_{\text{F}}+H_{\text{soc}}+H_{\text{I}},  \label{TH}
\end{equation}%
where the single atom Hamiltonian $H_{\text{F}}=\sum_{\mathbf{k},\sigma
=\uparrow ,\downarrow }(\epsilon _{\mathbf{k}}-\mu _{\sigma })C_{\mathbf{k}%
\sigma }^{\dagger }C_{\mathbf{k}\sigma }$, $C_{\mathbf{k}\sigma }^{\dagger }$
is the creation operator for a fermion atom with momentum $\mathbf{k}$ and
spin $\sigma $, $\epsilon _{\mathbf{k}}=k^{2}/2m$, $m$ is the atom mass, $%
\mu _{\uparrow }=\mu +h$, $\mu _{\downarrow }=\mu -h$, $\mu $ is the
chemical potential, and $h$ is the Zeeman field. The Hamiltonian for the
Rashba type of SO coupling is $H_{\text{soc}}=\alpha \sum_{\mathbf{k}%
}[(k_{y}-ik_{x})C_{\mathbf{k}\uparrow }^{\dagger }C_{\mathbf{k}\downarrow
}+(k_{y}+ik_{x})C_{\mathbf{k}\downarrow }^{\dagger }C_{\mathbf{k}\uparrow }]$%
. The interaction between atoms is described by $H_{\text{I}}=-g\sum_{%
\mathbf{k}}C_{-\mathbf{k}\uparrow }^{\dagger }C_{\mathbf{k}\downarrow
}^{\dagger }C_{\mathbf{k}\downarrow }C_{-\mathbf{k}\uparrow }$, where the
effective regularized interaction parameter $1/g=$ $\sum_{\mathbf{k}%
}1/(2\epsilon _{\mathbf{k}}+E_{b})$ for a 2D Fermi gas \cite{Chen,Liu,
Iskin2, He}. In experiments, the binding energy $E_{b}$ can be controlled by
tuning the $s$-wave scattering length or the barrier height $V_{0}$ along
the $z$ direction. Small and large $E_{b}$ correspond to the BCS and BEC
limit, respectively \cite{MR}.

The finite temperature properties of the 2D Fermi gas are obtained using
finite temperature quantum field theory, where the action for the
Hamiltonian (\ref{TH}) is $S_{V}=\int_{0}^{\beta }d\tau \lbrack \sum_{%
\mathbf{k},\sigma }C_{\mathbf{k}\sigma }^{\dagger }\partial _{\tau }C_{%
\mathbf{k}\sigma }+H]$ with $\beta =1/T$. Introducing the standard
Hubbard-Stratonovich transformation with the mean field superfluid order
parameter $\phi =g\sum_{\mathbf{k}}\left\langle C_{\mathbf{k}\downarrow }C_{-%
\mathbf{k}\uparrow }\right\rangle $ and integrating out the fermion degrees
of freedom, we have the partition function $Z=\int D\phi D\phi ^{\ast }\exp
(-S_{\text{eff}})$ with the effective action $S_{\text{eff}}=\int_{0}^{\beta
}d\tau (g^{-1}\left\vert \phi \right\vert ^{2}+\zeta _{\mathbf{k}})-\frac{1}{%
2}\mathbf{Tr}[\ln G^{-1}]$. Here
\begin{equation*}
G^{-1}=\left(
\begin{array}{cccc}
\partial _{\tau }+\zeta _{-,\mathbf{k}} & -\alpha k_{-} & 0 & -\phi  \\
-\alpha k_{+} & \partial _{\tau }+\zeta _{+,\mathbf{k}} & \phi  & 0 \\
0 & \phi ^{\ast } & \partial _{\tau }-\zeta _{-,\mathbf{k}} & -\alpha k_{+}
\\
-\phi ^{\ast } & 0 & -\alpha k_{-} & \partial _{\tau }-\zeta _{+,\mathbf{k}}%
\end{array}%
\right) ,
\end{equation*}
is the inverse Nambu matrix under the Nambu basis $\Psi (\mathbf{k})=(C_{%
\mathbf{k}\uparrow },C_{\mathbf{k}\downarrow },C_{-\mathbf{k\downarrow }%
}^{\dagger },C_{-\mathbf{k}\uparrow }^{\dagger })^{T}$, the symbol $\mathbf{%
Tr}$ denotes the trace over momentum, imaginary time, and the Nambu indices,%
\textbf{\ }$k_{\pm }=$\ $(k_{y}\pm ik_{x})$\textbf{\ }and $\zeta _{\pm ,%
\mathbf{k}}=(\epsilon _{\mathbf{k}}-\mu )\pm h=\zeta _{\mathbf{k}}\pm h$.
The superfluid order parameter $\Delta $ is obtained from the saddle point
of the effective action (i.e., $\frac{\partial S_{\text{eff}}}{\partial \phi
^{\ast }}|_{\phi =\Delta }=0$), which yields the gap equation
\begin{equation}
\sum_{\mathbf{k}}\frac{1}{(2\epsilon _{\mathbf{k}}+E_{b})}=\frac{1}{2}\sum_{%
\mathbf{k,}l=\pm }\left[ \lambda _{l}E_{l,\mathbf{k}}^{-1}\tanh (\beta E_{l,%
\mathbf{k}}/2)\right] .  \label{GP}
\end{equation}%
Here the quasiparticle energy spectrum $E_{\pm ,\mathbf{k}}=\sqrt{E_{\mathbf{%
k}}^{2}+\alpha ^{2}k^{2}+h^{2}\pm 2A}$, $A=\sqrt{\alpha ^{2}\zeta _{\mathbf{k%
}}^{2}k^{2}+h^{2}E_{\mathbf{k}}^{2}}$, $\lambda _{\pm }=\frac{1}{2}(1\pm
h^{2}/A)$, $E_{\mathbf{k}}^{2}=\zeta _{\mathbf{k}}^{2}+\left\vert \Delta
\right\vert ^{2}$, and $k^{2}=k_{x}^{2}+k_{y}^{2}$.

In 2D Fermi gases, it is well known that long-range superfluid order can be
destroyed by phase fluctuations of the order parameter at any finite
temperature \cite{PCH}. To study the phase fluctuations in SO coupled Fermi
gases, we set $\phi =\Delta e^{i\theta }$ following the standard procedure,
where $\theta $ is the superfluid phase around the saddle point (determined
by Eq. (\ref{GP})) that varies slowly in position and time spaces. The
superfluid phase can be decoupled from the original Green's function through
a unitary transformation $UG^{-1}(\theta )U^{\dagger }=G_{0}^{-1}-\Sigma $,
where $U=\exp (iM\theta /2)$, and $M=$ diag$(1,1,-1,-1)$. $\Sigma =\tau _{3}(%
{\frac{i\partial _{\tau }\theta }{2}}+{\frac{(\nabla \theta )^{2}}{8m}})-I({%
\frac{i\nabla ^{2}\theta }{4m}}+{\frac{i\nabla \theta \cdot \nabla }{2m}})+{%
\frac{\alpha }{2}}(\tau _{3}\sigma _{x}\partial _{y}\theta -I\sigma
_{y}\partial _{x}\theta )$ is the corresponding self-energy, $\sigma _{i}$
and $\tau _{i}$ are Pauli matrices in the Nambu space. $G_{0}^{-1}$ is the
Green's function at $\phi =\Delta $. The effective action can be written as $%
S_{\text{eff}}=S_{\text{0}}(\Delta )+S_{\text{fluc}}(\triangledown \theta
,\partial _{\tau }\theta )$ \cite{IAN}, where $S_{\text{fluc}}(\triangledown
\theta ,\partial _{\tau }\theta )=\text{tr}\sum_{n\geq 1}{\frac{1}{n}}%
(G_{0}\Sigma )^{n}$. Expanding $\Sigma $ up to the leading order ($n=2$), we
have
\begin{equation}
S_{\text{fluc}}=\frac{1}{2}\int d^{2}\mathbf{r}\left[ J(\triangledown \theta
)^{2}+P(\partial _{\tau }\theta )^{2}-Q(i\partial _{\tau }\theta )\right]
\label{SWA}
\end{equation}%
where $J={\frac{1}{4m}}\sum_{\mathbf{k}}(n_{\mathbf{k}}-\sum_{l=\pm }(%
\mathcal{J}_{1,l}+\mathcal{J}_{2,l}))$ represents the phase stiffness or
superfluid density. $n_{\mathbf{k}}=1-\sum_{l=\pm }{\frac{\xi (1+l\eta )}{%
2E_{l,\mathbf{k}}}}\text{Tanh}({\frac{\beta E_{l,\mathbf{k}}}{2}})$, $%
\mathcal{J}_{1,l}={\frac{m\alpha ^{2}}{2E_{l,\mathbf{k}}}}[(1-{\frac{%
k^{2}\alpha ^{2}\xi ^{2}}{2A^{2}}})+l({\frac{\Delta ^{2}+\xi ^{2}}{2A}}+{%
\frac{h^{2}((\Delta ^{2}+\xi ^{2})^{2}+k^{2}\alpha ^{2}\Delta ^{2}}{2A^{3}}}]%
\text{Tanh}({\frac{\beta E_{l,\mathbf{k}}}{2}})$, $\mathcal{J}_{2,l}={\frac{%
k^{2}\beta }{32m^{2}}}(1+l{\frac{m\alpha ^{2}\xi }{A}})^{2}\text{Sech}^{2}({%
\frac{\beta E_{l,\mathbf{k}}}{2}})$, $\eta =(h^{2}+k^{2}\alpha ^{2})/A$, $%
P=\sum_{\mathbf{k},l=\pm }{\frac{-\xi ^{2}(\eta +l)^{2}+E_{l,\mathbf{k}%
}^{2}(1+lh^{2}\Delta ^{2}\eta /A)}{8E_{l,\mathbf{k}}^{3}}}\text{Sech}({\frac{%
\beta E_{l,\mathbf{k}}}{2}})+{\frac{\beta \xi ^{2}(\eta +l)^{2}}{16E_{l,%
\mathbf{k}}^{2}}}\text{Sech}^{2}({\frac{\beta E_{l,\mathbf{k}}}{2}})$, and $%
Q=\sum_{\mathbf{k}}n_{\mathbf{k}}$. We have checked that the expressions for
$J$, $P$ and $Q$ reduce to previous results in different limits \cite%
{He,SSB,WZH,JTM}.

The phase $\theta $ can be decomposed into a static vortex part $\theta _{%
\text{v}}(\mathbf{r})$ and a time-dependent spin-wave part $\theta _{\text{sw%
}}(\mathbf{r},\tau )$. As a consequence, the action of the phase fluctuation
becomes $S_{\text{fluc}}=S_{\text{v}}+S_{\text{sw}}$ with $S_{\text{v}}=%
\frac{1}{2}\int d^{2}\mathbf{r}J[\triangledown \theta (\mathbf{r})]^{2}$ and
$S_{\text{sw}}=\frac{1}{2}\int d^{2}\mathbf{r}\{J[\triangledown \theta _{%
\text{sw}}(\mathbf{r},\tau )]^{2}+P[\partial _{\tau }\theta _{\text{sw}}(%
\mathbf{r},\tau )]^{2}-Q[i\partial _{\tau }\theta _{\text{sw}}(\mathbf{r}%
,\tau )]\}=\sum_{\mathbf{k}}\ln [1-\exp (-\beta \omega _{\mathbf{k}})]$,
where $\omega _{\mathbf{k}}=c\left\vert \mathbf{k}\right\vert $, $c=\sqrt{J/P%
}$ is the speed of the spin wave \cite{SSB}. Note that Eq. (\ref{SWA}) is
\textit{exactly} the same as the effective action for the 2D Heisenberg XY
model \cite{VLB,KT} with different $J$, $P$ and $Q$, therefore the BKT
transition temperature
\begin{equation}
T_{\text{BKT}}=\frac{\pi }{2}J(\Delta ,\mu ,T_{\text{BKT}}).  \label{BKT}
\end{equation}

Across the BKT temperature, there is a transition from the pseudogap phase
(with finite pairing $\Delta $ but without phase coherence or
superfludility) to phase coherent V-AV pairs (with both pairing and
superfluidity). When the temperature is further lowered, free V-AV pairs
form a tightly bounded V-AV lattice below another critical temperature \cite%
{DRN,APY},
\begin{equation}
T_{\text{Vortex}}={0.3}J(\Delta ,\mu ,T_{\text{Vortex}}).  \label{MT}
\end{equation}

The atom density equation can be obtained from the total thermodynamic
potential $\Omega =TS_{\text{eff}}$, yielding \cite{PM},
\begin{equation}
n=-\partial \Omega /\partial \mu =\sum\nolimits_{\mathbf{k}}n_{\mathbf{k}%
}-\beta ^{-1}\partial S_{\text{sw}}/\partial \mu ,  \label{NEQ}
\end{equation}%
where the atom density $n=mE_{\text{F}}/\pi $, and $E_{F}=\hbar
^{2}K_{F}^{2}/2m$ is the Fermi energy without SO coupling and Zeeman field.
The length unit is chosen as the inverse of the Fermi vector $K_{F}^{-1}$.

\begin{figure}[t]
\includegraphics[width=8.5cm]{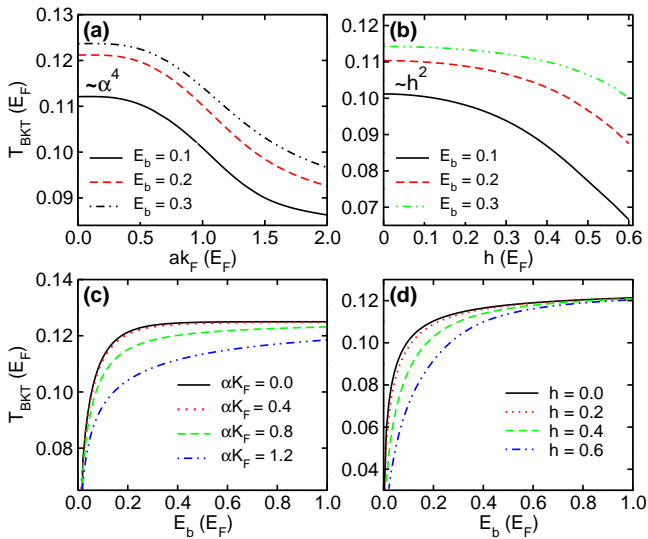}
\caption{(Color online) The dependence of the BKT temperature on SO coupling
(a), Zeeman field (b), and binding energy (c,d). $h=0$ in (a) and (c). $%
\protect\alpha K_{F}=1.0E_{F}$ in (b) and (d).}
\label{fig-BKT}
\end{figure}
We numerically solve Eqs. (\ref{GP}), (\ref{BKT}) and (\ref{NEQ})
self-consistently, and calculate various physical quantities. In Fig. \ref%
{fig-BKT}, we plot $T_{\text{BKT}}$ with respect to the SO coupling strength
$\alpha K_{F}$, the Zeeman field $h$, and the binding energy $E_{b}$. We
find that, quite surprisingly, $T_{\text{BKT}}$ decreases with increasing $%
\alpha $, although the superfluid order parameter $\Delta $ is enhanced \cite%
{Chen}. The unexpected decrease of $T_{\text{BKT}}$ has not been found in
previous literature \cite{He} and can be understood as follows. For $h=0$
and $\alpha K_{F}\ll 1$, the superfluid density
\begin{equation}
J(\alpha )-J(\alpha =0)\sim -\sum_{\mathbf{k}}\left[ {\frac{\Delta
^{2}k^{2}\alpha ^{4}}{8E_{0}^{5}}}+{\frac{e^{-\beta E_{0}}\beta \alpha ^{2}}{%
8}}\right] ,  \label{diff}
\end{equation}%
decreases with increasing $\alpha $. Here $E_{0}=\sqrt{\xi _{\mathbf{k}%
}^{2}+\Delta ^{2}}$. Physically, the increased density of states near the
Fermi surface dominates at small $\alpha $, hence enhances the phase
fluctuation and reduces the superfluid density. Note that the next leading
term in (\ref{diff}) is $\sim \alpha ^{6}$ for $T=0$ and $\sim \alpha ^{4}$
for $T\neq 0$, therefore the above analytical result is still valid even for
$\alpha K_{F}\sim E_{F}$. The Zeeman field is detrimental to the superfluid
order parameters, thus further reduces the superfluid density and the BKT
temperature, as shown in Fig. \ref{fig-BKT}b and Fig. \ref{fig-BKT}d.
Perturbation theory near $h\ll 1$ shows that the change of $T_{\text{BKT}}$
is $\sim h^{2}$, with the coefficient depending strongly on $\alpha $.
Similar features are also found for $T_{\text{Vortex}}$.

\begin{figure}[b]
\centering
\includegraphics[width=0.98\linewidth]{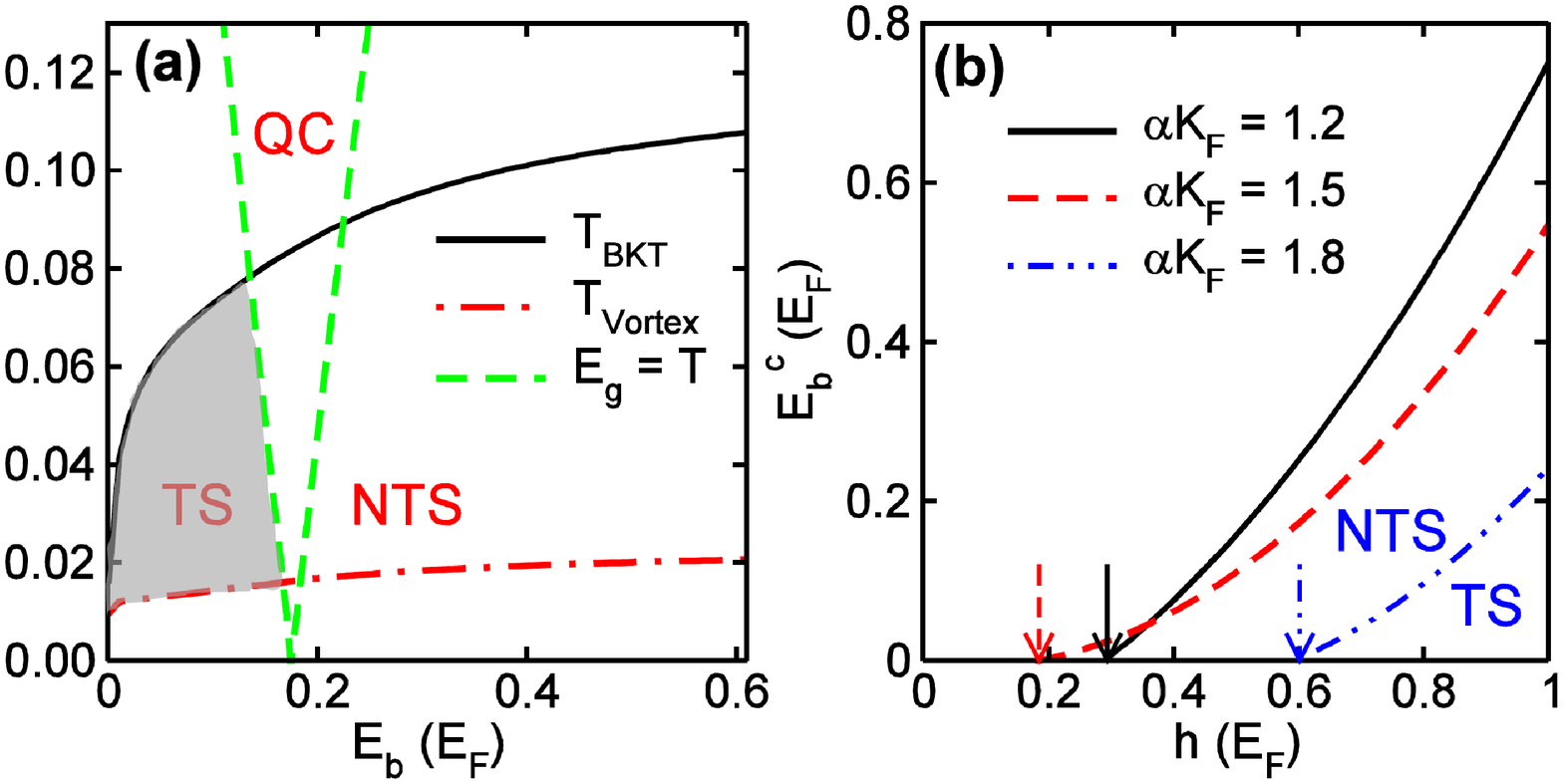} \vspace{-0.5cm}
\caption{(Color online). Parameter region for topological superfluids. (a)
QC: quantum critical region; TS: topological superfluids; NTS:
non-topological superfluids. The shadow region is the possible parameter
region for observing MFs. $h=0.6E_{F}$, $\protect\alpha K_{F}=1.5E_{F}$. (b)
Phase boundary between TS and NTS at $T=0$. The arrows mark the required
minimum Zeeman fields.}
\label{fig-phase}
\end{figure}

In the presence of both SO coupling and Zeeman field, a topological
superfluid can emerge from regular $s$-wave interaction when $h$ is larger
than a critical value $h_{c}=\sqrt{\mu ^{2}+\Delta ^{2}}$ \cite{JDS,GM}.
Around $h_{c}$, the minimum quasiparticle energy gap occurs at $k=0 $ (i.e.,
$E_{g}=E_{-,k=0}$), which first closes and then reopens across $h_{c}$,
allowing the system to change its topological order from a regular \textit{s}%
-wave superfluid to a topological superfluid where MFs exist in vortex cores
\cite{JDS}. At finite temperature, the zero temperature topological phase
transition becomes a phase crossover. In Fig. \ref{fig-phase}a, we plot the
line $E_{g}=T$ for a finite $h$, which is obtained by solving Eqs. (\ref{GP}%
), (\ref{NEQ}) and $E_{g}=T$ self-consistently. The gap closes at a critical
$E_{b}^{c}$ where $h_{c}=\sqrt{\mu ^{2}+\Delta ^{2}}=h$. When $E_{b}<$ ($>$)
$E_{b}^{c}$, $h>$ ($<$) $h_{c}$, and the region below the line $E_{g}=T$
corresponds to the topological (non-topological) superfluid. Above the line $%
E_{g}=T$, the temperature is larger than the quasiparticle energy gap and
the thermal excitations destroy the topological superfluid. We emphasize
that at finite temperature there is no sharp phase transition, but only
phase crossover between topological and non-topological superfluids. The
line $E_{g}=T$ is plotted only for guiding purpose and there is no sharp
boundary between different phases.

The solid and dash dotted lines in Fig. \ref{fig-phase}a correspond to $T_{%
\text{BKT}}$ and $T_{\text{Vortex}}$, respectively. We see $T_{\text{Vortex}%
} $ quickly approaches a constant $T_{\text{Vortex}}=3E_{F}/40\pi $ with
increasing binding energy. In the pseudogap region ($T>T_{\text{BKT}}$),
free vortices may exist, but they are not suitable for the observation of
MFs due to the lack of phase coherence. While in the vortex lattice region ($%
T<T_{\text{Vortex}}$), the zero energy modes may break into two normal
states with energy splitting $\propto \Delta e^{-R/\xi }$ \cite%
{Cheng,Kitaev2} because of the large tunneling between neighboring vortices
in the lattice, where $\xi $ is the coherence length of the superfluid, and $%
R$ is the intervortex distance. In this region, strong disorder in the
tunneling may also lead to Majorana metals \cite{disorder}. Clearly, the
required temperature for observing MFs should be $T_{\text{Vortex}}<T<T_{%
\text{BKT}}$, instead of the intuitive zero temperature \cite{Liu,Iskin2}.
In Fig. \ref{fig-phase}a, the shadow regime between $T_{\text{BKT}}$ and $T_{%
\text{Vortex}}$ gives the possible parameter range for the experimental
observation of MFs which exist in the cores of the naturally present V-AV
pairs in this region. Note that the intervortex distance $R$ is still
essential for the observation of MFs in this region. Such distance may be
estimated using an analogy between V-AV pairs and the 2D Coulomb gases. It
has been shown \cite{leggett} that the mean-square radius $\langle
R^{2}\rangle \sim \xi ^{2}\frac{\pi \beta J-1}{\pi \beta J-2}$, which
diverges at $T=T_{\text{BKT}}$ (see Eq. (\ref{BKT})) and the V-AV pair
breaks into free vortices. Therefore there should exist a finite temperature
regime below $T_{\text{BKT}}$ where $R\gg \xi $ and the splitting of the
zero energy Majorana states is vanishingly small. Note here that our theory
can only capture the average behavior of V-AV pairs and a more delicate
theory is still needed to further understand the detailed structure of V-AV
pairs.

In Fig. \ref{fig-phase}b, we plot the parameter region for topological
superfluids with respect to the Zeeman field and the binding energy at the
zero temperature. Generally, the required critical $E_{b}^{c}$ for
topological superfluids increases when $h$ increases. Note that here the
phase boundary is determined by $h_{c}=\sqrt{\mu ^{2}+\Delta ^{2}}$ at $T=0$%
, but does not shift much even at finite temperature.

\begin{figure}[t]
\includegraphics[width=1.8in]{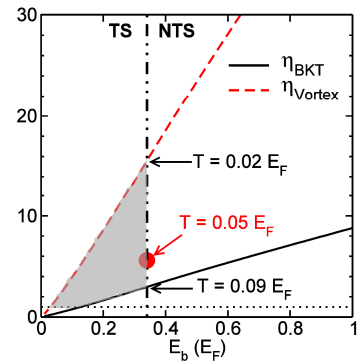}
\caption{(Color online) Plot of $\protect\eta =\protect\varepsilon _{m}/T$
as a function of $E_{b}$. $h=0.8E_{F}$ and $\protect\alpha K_{F}=1.6E_{F}$.
Dashed-Dotted line: $E_{b}^{c}=0.33E_{F}$ as the corresponding boundary
between topological superfluids (TS) and non-topological superfluids (NTS)
(see the $E_{g}=T$ line in Fig. \protect\ref{fig-phase}a for the
determination of the phase boundary). Dotted line: $\protect\eta =1$. Filled
circle corresponds to $\protect\eta $ at $T=0.05E_{F}$. The shadow region is
the possible parameter region for observing MFs. }
\label{fig-eta}
\end{figure}

Because MFs can only be observed at finite temperature, there exists a
nonzero probability $\sim \exp \left( -\eta \right) $ for thermal
excitations to non-topological excited states in the vortex core, where the
ratio $\eta =\epsilon _{m}/T$, $\epsilon _{m}\sim \Delta ^{2}/(2E_{F})$ is
the minimum energy gap (minigap) \cite{Ho, Mao} in the vortex core that
protects zero energy MFs. For the observation of MFs, it is crucially
important to have $\eta >1$, in addition to the requirement of the
temperature $T_{\text{Vortex}}<T<T_{\text{BKT}}$ (i.e., $\eta >\eta _{\text{%
BKT}}\equiv \epsilon _{m}/T_{\text{BKT}}$ and $\eta <\eta _{\text{Vortex}%
}\equiv \epsilon _{m}/T_{\text{Vortex}}$). When $E_{b}\gg E_{F}$, we have $%
\eta _{\text{BKT}}=8E_{b}/E_{F}$ and $\eta _{\text{Vortex}}=40\pi
E_{b}/E_{F} $, which means $\eta \gg 1$ for a large $E_{b}$. Remarkably, $%
\eta $ is also dramatically enhanced by the SO coupling in the BCS side. In
Fig. \ref{fig-eta}, we plot $\eta _{\text{BKT}}$ and $\eta _{\text{Vortex}}$
with respect to $E_{b}$ for parameters $h=0.8E_{F}$ and $\alpha
K_{F}=1.6E_{F}$. The vertical line at $E_{b}\simeq 0.33E_{F}$ is the
boundary between topological and non-topological superfluids. There is a
broad region in the BCS side with $\eta _{\text{BKT}}>1$ even at the highest
temperature $T_{\text{BKT}}$. The region can be much larger when the
temperature is further lowered to $T=T_{\text{Vortex}}$. The solid circle
represents the possible temperature $T=0.05E_{F}$ that may be accessible in
experiments in the near future \cite{MF,Jin, NoteJin}, yielding $\eta \sim
5.6$ and the thermal excitation probability less than 0.4\%. Such a small
probability clearly demonstrates that MFs can actually be observed with
realistic temperature in experiments. Note that in the presence of SO
coupling $\epsilon _{m}$ may be larger than $\Delta ^{2}/(2E_{F})$ used for
our estimate \cite{Mao}, therefore $\eta $ could be even larger, which
further suppresses thermal excitations.

We illustrate how to find suitable parameters, in particular $E_{b}$, for
the observation of MFs. Because topological superfluids only exist in the
region $h>\sqrt{\mu ^{2}+\Delta ^{2}}$, we need $|\Delta |^{2}\gg |\mu |^{2}$
to obtain a small $h$ and a large $\Delta $. Clearly the BEC limit with
large $E_b$ does not work because $|\mu |\sim |E_{F}-E_{b}/2|\gg \Delta \sim
\sqrt{2E_{b}E_{F}}$. While in the BCS side, $|\Delta |^{2}\gg |\mu |^{2}$
can be achieved by choosing suitable SO coupling, Zeeman field and binding
energy \cite{Chen}. To obtain a large quasiparticle minigap in the vortex
core (thus a large $\eta $) for MFs, the binding energy should be set to be
close to the critical $E_{b}^{c}$ (see Fig. \ref{fig-eta}). At the same
time, the bulk quasiparticle gap should also be chosen to be much larger
than the temperature. Note that in experiments the bulk quasiparticle gap
for the topological superfluid can be detected using the recently
experimentally demonstrated momentum-resolved photoemission spectroscopy
\cite{MRPL}.

Finally we briefly compare the cold atomic gases \cite{Zhang} with the
semiconductor-superconductor nanostructure where certain signature of MFs
has been observed in experiments \cite{mf1,mf2,mf3,mf4}. Both systems share
the same ingredients for MFs: SO coupling, Zeeman fields and s-wave pairing,
besides the \textit{s}-wave pairing is through intrinsic \textit{s}-wave
interaction in cold atoms, while externally induced in the nanostructure.
Because of high controllability and free of disorder (lacked in the
corresponding solid state systems), the topological cold atomic superfluids
provide an ideal and promising platform for observing MFs and the associated
non-Abelian statistics, which are both fundamentally and technologically
important.

In summary, we study the BKT transition in 2D SO coupled Fermi superfluids
and find the unexpected decrease of the BKT transition and vortex lattice
melting temperatures with increasing SO coupling. We characterize the finite
temperature phase diagram for the experimental observation of MFs in this
system. Our work not only provides the basis for future study of rich and
exotic 2D SO coupled Fermi superfluid physics, but also yields realistic
parameter regions for experimentally realizing nontrivial topological
superfluid states from stable cold atom \textit{s}-wave superfluids.

We thank Yongping Zhang, Li Mao and Lianyi He for helpful discussions. This
work is supported partly by DARPA-YFA (N66001-10-1-4025), ARO
(W911NF-09-1-0248), NSF (PHY-1104546), AFOSR (FA9550-11-1-0313), and
DARPA-MTO (FA9550-10-1-0497). G.C. and S.J. are also supported by the 973
program under Grant No. 2012CB921603, the NNSFC under Grant Nos. 10934004,
60978018, and 11074154. M.G. and G.C. contributed equally to this work.


\begin{thebibliography}{99}
\bibitem{NoteJin} $T=0.05E_{F}$ has been realized for 3D Fermi gases \cite%
{Jin}, while $T\sim 0.2E_{F}$ has been achieved for 2D Fermi gases \cite{MF}%
. We expect that $T=0.05E_{F}$ should be achievable for 2D Fermi gases in
the near future (private communication with M. K\"{o}hl).
\end{thebibliography}
\end{document}